\documentclass[copyright,creativecommons]{eptcs}

\usepackage{enumitem}
\usepackage{xspace}
\usepackage{graphicx}
\usepackage[english]{babel}
\usepackage[utf8]{inputenc}
\usepackage{times}
\usepackage[T1]{fontenc}
\usepackage{lipsum}
\usepackage{inconsolata}
\usepackage{ragged2e}
\usepackage{alltt}
\usepackage{rotating}
\usepackage{lscape}
\usepackage{setspace}
\usepackage{array}
\usepackage{verbatim}
\usepackage{fancyvrb}
\usepackage{upquote}
\usepackage[dvipsnames]{xcolor}
\usepackage{tikz}
\usetikzlibrary{arrows}
\usepackage[htt]{hyphenat}
\PassOptionsToPackage{hyphens}{url}
\usepackage{hyperref}

\allowhyphens

\newcommand{\secref}[1]{Section~\ref{#1}}

\newcommand{\code}[1]{\texttt{#1}}
\newenvironment{codeblock}{\begin{alltt}}{\end{alltt}}

\newcommand{\acldoc}[2] 
 {\cite[\href{#2}{\texttt{:doc} \underline{\texttt{#1}}}]{acl2-manual}}

\begin{document}

\title{Adding 32-bit Mode to the ACL2 Model of the x86 ISA}

\author{Alessandro Coglio
        \institute{Kestrel Technology LLC, Palo Alto, CA (USA)}
        \email{coglio@kestreltechnology.com}
        \and
        Shilpi Goel
        \institute{Centaur Technology, Inc., Austin, TX (USA)}
        \email{shilpi@centtech.com}}

\def\titlerunning{32-bit Mode of the x86 ISA}
\def\authorrunning{Coglio and Goel}

\maketitle

\begin{abstract}
The ACL2 model of the x86 Instruction Set Architecture
was built for the 64-bit mode of operation of the processor.
This paper reports on our work to extend the model
with support for 32-bit mode,
recounting the salient aspects of this activity
and identifying the ones that required the most work.
\end{abstract}


\section{Motivation and Contributions}\label{sec:introduction}

A formal model of the ISA (Instruction Set Architecture)
of the pervasive x86 processor family
can be used to
clarify and disambiguate its informal documentation~\cite{intel-manuals},
as well as support
the verification and formal analysis of existing binary programs
(including malware),
the verified compilation of higher-level languages to machine code,
the synthesis of correct-by-construction binary programs
from specifications,
and the verification that micro-architectures correctly implement the ISA.
Furthermore, if the model is efficiently executable,
it can be used as a high-assurance simulator, e.g.\ as a test oracle.

The ACL2 model of the x86 ISA
\acldoc{x86isa}{http://www.cs.utexas.edu/users/moore/acl2/manuals/current/manual/?topic=ACL2____X86ISA}~\cite{goel-dissertation} is one such model.
It includes a substantial number of instructions and
has been used to verify several non-trivial binary programs.
Its fast execution has enabled its validation
against actual x86 processors on a large number of tests.

Prior to the work described in this paper,
the model supported the simulation and analysis of 64-bit software only
(which includes modern operating systems and applications),
and was used to verify several 64-bit programs.
We refer to this pre-existing model of the x86 ISA as the \emph{64-bit model}.

However, legacy 32-bit software is still relevant.
Many applications are still 32-bit,
running alongside 64-bit applications in 64-bit operating systems.
Most malware is 32-bit~\cite{malware64bits};
as part of a project to detect malware variants
via semantic equivalence checking by symbolic execution,
we were given a large corpus of known Windows malware for experimentation,
and found indeed that it mainly contained 32-bit executables.\footnote{This statement
  refers to a project at Kestrel Technology.}
Verifying the 32-bit portion of a micro-architecture
requires a model of the 32-bit ISA.
Also, the aforementioned
verified compilation of higher-level languages to machine code
and synthesis of correct-by-construction binary programs from specifications,
while presumably generally oriented towards 64-bit code,
may occasionally need to target 32-bit code,
e.g., to interoperate with legacy systems.

Thus, we worked on extending the 64-bit model with 32-bit support:
all the non-floating-point instructions in the 64-bit model
have been extended from 64 bits to 32 bits;
furthermore, a few 32-bit-only instructions have been added.
However,
the 32-bit extensions have not been validated against actual x86 processors yet
as thoroughly as the 64-bit portions of the model.
Work on verifying 32-bit programs using this model has also started
but is too preliminary to report.
We refer to this extended model of the x86 ISA as the \emph{extended model}.

This paper reports on
our work to extend the 64-bit model.
\secref{sec:background} provides
some background on the x86 ISA and on the 64-bit model.
\secref{sec:extension} recounts
the salient aspects of the 32-bit extensions to the model,
including how they were carried out
and which ones required the most work.
Related and future work are discussed in
\secref{sec:related} and \secref{sec:future}.
The 64-bit model was developed by the second author~\cite{goel-dissertation},
and the 32-bit extensions were developed by the first author
(with some help and advice from the second author);
this suggests that the model is extensible by third parties.


\section{Background}\label{sec:background}

We present a very brief overview of the x86 ISA that is relevant to
this paper in Section~\ref{sec:background-x86}.  In
Section~\ref{sec:background-model}, we describe some features of the
64-bit model of the x86 ISA.

\subsection{x86 ISA: A Brief Overview}\label{sec:background-x86}

Intel's modern x86 processors offer various modes of
operation~\cite[Volume 1, Section 3.1]{intel-manuals}.
The IA-32 architecture supports 32-bit
computing by giving access to $2^{32}$ bytes of 
memory, and it
provides three modes:
\begin{enumerate}[nosep]
\item {\it Real-address mode}, which is the x86 processor's mode upon
  power-up or reset.
\item {\it Protected mode}, which is informally referred to as the
  ``32-bit mode''; this mode can also emulate the real-address mode if
  its {\it virtual-8086 mode} attribute is set.
\item {\it System management mode}, which is used to run firmware to
  obtain fine-grained control over system resources to perform
  operations like power management.
\end{enumerate}
The Intel 64 architecture added the {\em IA-32e mode}, which has two
sub-modes:
\begin{enumerate}[nosep]
\item {\em 64-bit mode}, which gives access to $2^{64}$ bytes of 
  memory, and thus, allows the execution of 64-bit code.
 \item {\em Compatibility mode}, which allows the execution of legacy
   16- and 32-bit code inside a 64-bit operating system; in this
   sense, it is similar to 32-bit protected mode vis-\`{a}-vis
   application programs.
\end{enumerate}
The 64-bit model of the x86 ISA, discussed briefly in
Section~\ref{sec:background-model}, specified only the 64-bit sub-mode
of IA-32e mode.  In the subsequent sections, we focus on 
32-bit protected mode and compatibility mode.  Modeling the rest of the
modes is a task for the future (see Section~\ref{sec:future}).

IA-32e and 32-bit protected modes are the most used operating
modes of x86 processors today. A big difference between these modes is
in their memory models --- we first briefly discuss the memory
organization on x86 machines.  There are two main types of memory
address spaces: {\em physical address space} and {\em linear address
  space}.  The physical address space refers to the processor's main
memory; it is the range of addresses that a processor can generate on
its address bus.  The linear address space refers to the processor's
linear memory, which is an abstraction of the main memory
(see paging below).  Programs
usually do not access the physical address space directly; instead,
they access the linear address space.

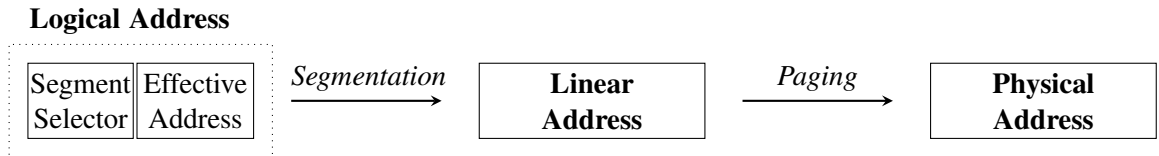
\begin{figure}[h]
  \centering
  \begin{tikzpicture}
    \draw (0,0) rectangle (1.4,1);  \node[text width=1.5cm,align=center] at (0.7, 0.5) {Segment Selector};
    \draw (1.45,0) rectangle (3,1);  \node[text width=1.5cm,align=center] at (2.225, 0.5) {Effective Address};

    \draw[dotted] (-0.25,-0.25) rectangle (3.25,1.25);  \node[text width=3cm,align=center,above] at (1.35, 1.25) {{\bf Logical Address}};
    \node[text width=1.5cm,align=center,above] at (4.25, 0.5) {{\em Segmentation}};
    \draw [>=stealth,->,thick] (3.5,0.5) -- (5.5,0.5);
    \draw (6,0) rectangle (9,1);  \node[text width=1.5cm,align=center] at (7.5, 0.5) {{\bf Linear Address}};
    \node[text width=1.5cm,align=center,above] at (10.5, 0.5) {{\em Paging}};
    \draw [>=stealth,->,thick] (9.5,0.5) -- (11.5,0.5);
    \draw (12,0) rectangle (15,1); \node[text width=1.5cm,align=center] at (13.5, 0.5) {{\bf Physical Address}};
  \end{tikzpicture}
  \caption[Memory Addresses]{\small{Memory Addresses}}
  \label{fig:back-memory-addressing}
\end{figure}

{\em Segmentation} and {\em paging} are two main facilities on x86
machines that manage a program's view of the memory; see
Figure~\ref{fig:back-memory-addressing}.  When using segmentation,
programs traffic in {\em logical addresses}, which consist of a {\em
  segment selector} and an {\em effective address}.
The x86 processors provide six (user-level) segment selector registers ---
\code{CS}, \code{SS}, \code{DS}, \code{ES}, \code{FS}, and \code{GS}.
A segment selector
points to a data structure, called a \emph{segment descriptor}, that contains, among other information, the
{\em segment base}, which specifies the starting address of the segment,
and the {\em segment limit}, which specifies the size of the segment.  As an
optimization, the processor caches information from this data
structure in the {\em hidden} part of these registers whenever a
selector register is loaded.  The segment base is added to the
effective address to obtain the linear address.  The upshot of this
is that the linear address space is divided into a group of segments.
An operating system can
assign different sets of segments to different programs or even
different segments to the same program (e.g., separate ones for code,
data, and stack) to enforce protection and non-interference.  It can
also assign the same set of segments to different processes to enable
sharing and communication.  The processor makes three main kinds of
checks when using segmentation: (1) limit checks, which cause an
exception if code attempts to access memory locations outside the
segment; (2) type checks, which cause an exception if code uses a
segment in some unintended manner (e.g., an attempt is made to write to
a read-only data segment); and (3) privilege checks, which
cause an exception if code attempts to access a segment which it does
not have permission to access.

Paging is used to map linear addresses to physical addresses. It is
conventional to define this mapping in such a manner that a larger
linear address space can be simulated by a smaller physical address
space.  Each segment is divided into smaller pages, which can be
resident either in the physical memory (if the page is in use) or on
the disk (if the page is not currently in use).  An operating system
can swap pages back and forth from the physical memory and the disk;
thus, paging supports a {\em virtual} memory environment.  There are
three kinds of paging offered by x86 processors: 32-bit (translates
32-bit linear addresses to up to 40-bit physical addresses), PAE
(translates 32-bit linear addresses to up to 52-bit physical
addresses), and 4-level (translates 48-bit linear addresses to
up to 52-bit physical addresses).

In 64-bit mode, a logical address, an effective address, and a linear
address for a memory location are usually the same because all segment
bases, except those for two out of six segment selectors (i.e., \code{FS}
and \code{GS}), are treated as zero.  Moreover, limit checking and
a certain kind of type checking are not performed in this mode.  Thus,
segmentation is effectively a legacy feature in 64-bit mode.  However,
in 32-bit and compatibility modes, segmentation is used in its
full glory.

32-bit protected and IA-32e modes also differ with regard to
paging.  In 32-bit protected mode, 32-bit and PAE paging are used
to translate 32-bit linear addresses.  In both sub-modes of IA-32e
mode, only 4-level paging is used.  However, since 4-level paging
translates only 48-bit linear addresses and IA-32e mode has 64-bit
linear addresses, this mode has a notion of {\em canonical addresses}.
Canonical addresses are 64-bit linear addresses whose bits 63 through
47 are identical --- either all zeroes or all ones.  Additionally,
since compatibility mode traffics in 32-bit addresses, bits 47 through
32 of a 64-bit linear address are treated as zero.  Any attempt to map
a non-canonical linear address to a physical address in IA-32e
mode will lead to an exception.

Of course, there are many other differences between 32-bit 
and 64-bit modes.
Even instruction decoding differs in these modes.
An x86 instruction has a variable length up to 15 bytes:
the {\em opcode} bytes in the instruction determine which
operation is to be performed; the other instruction bytes act as size
modifiers, or specify the location of the operands, etc.  64-bit
instructions can have additional bytes preceding an opcode, called the
\code{REX} prefixes, which can modify the operation's sizes or
operands' locations.  In 32-bit mode, these bytes actually
correspond to the increment/decrement (\code{INC}/\code{DEC}) opcodes.
Task management is another big example --- a task can consist of a
program, a process, or an interrupt or exception handler, and it is
executed and/or suspended by the processor as an atomic unit.  Task
management is essentially the software's responsibility in 64-bit
mode, but it is handled by hardware in 32-bit protected mode.
Generally speaking, 32-bit mode and compatibility sub-mode
support many legacy features, and as such, they are less
``streamlined'' than the newer 64-bit mode.

We refer the reader to Intel's official
documentation~\cite{intel-manuals} for more information about the x86
ISA.

\subsection{\code{x86isa}: A Model of the 64-bit Mode}\label{sec:background-model}

Prior to this work, the \code{x86isa} library in the ACL2 community
books contained the specification of a significant subset of
64-bit sub-mode of IA-32e mode. This 64-bit model follows the
classical {\em interpreter style of operational
  semantics}~\cite{mccarthy-algol,mechanized-op-semantics} ---
it can be described by
the following main components:
\begin{itemize}[nosep]
\item a machine state consisting of registers, memory, flags, etc.;
\item instruction semantic functions that define the behavior of
  x86 instructions in terms of updates made to the machine state;
\item a step function that fetches an x86 instruction from the
  memory, decodes it till the opcode byte(s) are read, and then
  executes it by calling a top-level opcode dispatch function --- this
  dispatch function is basically a giant case statement that calls the
  appropriate instruction semantic function corresponding to the
  opcode; and
\item a run function, which is the top-level interpreter that calls
  the step function iteratively.
\end{itemize}

The 64-bit model offers two views of x86 machines: the {\em
  system-level} view --- when the \code{app-view} field in the machine
state is \code{nil}, and the {\em application-level} view --- when the
\code{app-view} field is not \code{nil}.  The system-level view is
intended to specify the x86 ISA as accurately as possible.  In this
view, one can verify system programs that have access to system state,
such as the memory management data structures.
The application-level view can be used to simulate and verify application
programs in the same sort of environment that programmers use to
develop such programs.  In this view, the operating assumption is that
the underlying operating system and system features of the ISA
behave as expected.
For instance, unlike the system-level view, this view does not include
the model of paging --- it offers the linear memory abstraction
instead of the physical memory.  Also, the application-level view does not
contain the specification of system instructions (which are present in
the system-level view), e.g., instruction \code{LGDT}, which
manipulates the segmentation data structures.  However, the application-level
view does specify 64-bit segmentation --- i.e.,
\code{FS}/\code{GS}-based logical to linear address translation,%
\footnote{Recall from
  Section~\ref{sec:background-x86} that for other segment selectors,
  the linear and logical addresses are the same in 64-bit mode.} --- but
here it makes the following assumption: the hidden parts of these
segment selectors are assumed to contain the correct segment base
addresses.  This assumption is reasonable because these hidden parts
of the registers are automatically populated by the processor when the
registers are loaded, and the instructions that load these registers
are system-level instructions --- thus, it is assumed that an operating system
routine correctly loaded these registers.  These views aim to mitigate
the trade-off between verification utility and verification effort.
Of course, if such assumptions that come with the application-level view's
higher-level of abstraction are unacceptable to a user, he or she can
operate in the system-level view for {\em all} programs.

This model has been used to do proofs about both application
and system programs.  Some examples of application programs verified
using the model are a population count program, a word-count program,
and a data-copy program.  All of these programs except population
count were verified by using the lemma libraries that are included in
the \code{x86isa} library; the population count program was verified
completely automatically using the \code{GL}
library~\cite{gl-diss,bit-blasting-GL}.  The 64-bit model is also
compatible with the Codewalker library~\cite{acl2:books:codewalker};
Codewalker has been used to reason about small application programs
like the x86 machine-code routine to compute the factorial of a
number.  An example of a system program verified using the \code{x86isa}
library is zero-copy, which efficiently copies data from a
source to a destination memory location by modifying the memory
management data structures (i.e., using the {\em copy-on-write}
technique).

The \code{x86isa} library uses ACL2 features like abstract
stobjs~\cite{abstract-stobjs} to provide both reasoning and execution
efficiency.  The simulation speed of the 64-bit model is around
300,000 instructions/second in the system-level view and 3 million
instructions/second in the application-level view.\footnote{All such measurements mentioned in this paper have been done on a
  machine with Intel Xeon E31280 CPU @ 3.50GHz with 32GB RAM.}  As a
direct consequence, it has been possible to extensively validate the
64-bit model by running co-simulations against Intel's x86 machines,
thereby increasing confidence in the model's accuracy.  A goal of the
\code{x86isa} library is to make the x86 ISA specification accessible
to users, and as such, these books are extensively documented
\acldoc{x86isa}{http://www.cs.utexas.edu/users/moore/acl2/manuals/current/manual/?topic=ACL2____X86ISA} --- both from a user and a developer's
point of view.

For more about the 64-bit model and its use in 64-bit code analysis,
we refer an interested reader elsewhere ---
see~\cite{DBLP:journals/corr/Goel17} for a more detailed overview
and~\cite{goel-dissertation,x86isa-procos} for an in-depth report.


\section{Extension of the Model}\label{sec:extension}

In this section, when we refer to code running in {\em
  32-bit mode}, we mean {\em application} software running in either
32-bit protected mode or compatibility sub-mode of IA-32e mode.

\subsection{Challenges}\label{sec:extension:challenges}

Extending the model to 32-bit mode was not simply a matter of
generalizing the sizes of the operands and addresses
manipulated by the instructions.
As explained in \secref{sec:background-x86},
32-bit mode is more complicated than 64-bit mode,
due to the legacy features that it provides.

In particular,
in 32-bit mode, memory accesses are more complicated than in 64-bit mode:
32-bit mode uses segmentation,
which is almost completely disabled in 64-bit mode.
The 64-bit model made, throughout, the reasonable assumption that
the effective addresses in
instruction pointer, stack pointer, addressing mode base registers, etc.\
were also linear addresses --- the only exception was when \code{FS} and \code{GS} segment selectors
were used, in which case their bases were explicitly added
to effective addresses to get the corresponding linear addresses.
Effective/linear addresses were checked to be canonical as needed.
This had to be generalized in the extended model:
effective and linear addresses had to be differentiated,
and segment bound checks had to be performed,
before adding segment bases to effective addresses.

This required some changes
in the interfaces between certain parts of the model,
e.g., in the signature of the ACL2 functions to read and write memory.
Generally, changes to interfaces and their semantics may break proofs
whose restoration may require different or more general lemmas and invariants.
As explained in \secref{sec:background-model},
the 64-bit model was accompanied
by several correctness proofs of non-trivial 64-bit programs ---
besides the termination, guard, and other proofs in the model proper.
To keep the adaptation of all these proofs more manageable,
the changes had to be carried out in incremental steps, to the extent possible.
An overarching goal was to keep the whole build working at all times.

\subsection{Mode Discrimination}\label{sec:extension:mode}

Since the processor does certain things differently
depending on whether the current mode is 64-bit or 32-bit,
a predicate is needed to discriminate between the two modes in the model.
The 64-bit model included a predicate \code{64-bit-modep} over the x86 state,
defined to be always \code{t}, which was rarely called.
This predicate was extended to distinguish between the two modes:
if the \code{LMA} bit of the \code{IA32\_EFER} register is set
(i.e., the processor is in IA-32e mode),
and the \code{L} bit of the \code{CS} register%
\footnote{By expressions like
`bit \textit{x} of the segment register \textit{y}'
we mean, more precisely,
`bit \textit{x} of the segment descriptor
whose selector is currently loaded in the segment register \textit{y}'.}
is set
(i.e., the current application is running in 64-bit sub-mode,
not in compatibility 32-bit sub-mode),
the predicate returns \code{t};
otherwise, the processor is in
either protected 32-bit mode
or compatibility 32-bit mode,
and the predicate returns \code{nil}.
Later, calls of this predicate were replaced by
\code{(eql proc-mode *64-bit-mode*)},
where the value of the variable \code{proc-mode} is the current processor mode,
read once from the x86 state at each execution step
and passed to many of the model's functions;
for simplicity, in the rest of the paper we still show
calls of \code{64-bit-modep}.

In order to extend the instructions covered by the 64-bit model
to 32-bit mode one by one,
each branch of the top-level opcode-based dispatch
was wrapped into a conditional of this form:
\begin{codeblock}
  (if (64-bit-modep x86) <do-as-before> <throw-error>)
\end{codeblock}
This let the model behave as before in 64-bit mode,
and provided a clear indication, in concrete or symbolic execution,
of instructions not yet extended to 32-bit mode.
The addition of these conditionals required
the addition of many \code{64-bit-modep} hypotheses to the existing theorems
to keep all the proofs working
(which rely on instructions not throwing errors in normal circumstances),
a tedious but simple task;
however, the extensions described in the subsequent sections required
additional work to keep all the proofs working ---
see \secref{sec:extension:proofs} for details.

\subsection{Segmentation}\label{sec:extension:segmentation}

Instructions manipulate operands in registers and memory,
including stack operands and immediate operands.
The register access functions in the 64-bit model
actually needed no extension for 32-bit mode.
However, as mentioned in \secref{sec:extension:challenges},
the memory access functions in the 64-bit model
had to be extended to take segmentation into account.
These memory access functions are used not only for operands,
but also for fetching the bytes that form instructions
(prefixes, opcodes, etc.).
So adding segmentation was a prerequisite to extending any instruction.
Segmentation was modeled in a form that
is used uniformly in both 64-bit and 32-bit mode,
as explained below.

The 64-bit model already included state components for the segment registers,
including their hidden parts.
A new function \code{segment-base-and-bound} was added
to retrieve the base and bounds of a segment
from (the hidden part of) the corresponding segment register in the x86 state.
This function takes as arguments
the machine state and (the identifying index of) a segment register
and returns the base address, lower bound, and upper bounds of the segment
as results via \code{mv}.
More precisely,
in 64-bit mode, the bases of the \code{FS} and \code{GS} segments
are retrieved from model-specific registers
that are physically mapped (in the processor) to
the hidden base fields of the \code{FS} and \code{GS} registers,
and that provide 64-bit linear addresses
beyond the 32-bit linear addresses used in 32-bit mode.
The bounds of a segment refer to effective addresses,
which are offsets into the segment.
The lower bound is 0
and the upper bound is the limit field in the segment register,
unless the \code{E} bit in the segment register is set.
This \code{E} bit indicates an expand-down segment (typically used for a stack):
the lower bound is the limit field in the segment register,
and the upper bound is either $2^{32}-1$ or $2^{16}-1$,
depending on the \code{D/B} bit in the segment register,
which indicates a 32-bit segment if set and a 16-bit segment if cleared.
In 64-bit mode, since the bounds are ignored,
\code{segment-base-and-bounds} just returns 0 as both bounds;
it also just returns 0 as the base
if the segment register is not \code{FS} or \code{GS}.

A new function \code{ea-to-la} was added
to perform segment address translation;
the name stands for `effective address to linear address',
chosen in analogy to the \code{la-to-pa} function that,
in the 64-bit model, performed paging address translation,
i.e., turned linear addresses into physical addresses.%
\footnote{\code{ea-to-la} actually translates
a logical address to a linear address,
but the name \code{la-to-la} would not have worked well.
There is a sense in which the effective address portion of a logical address
is more ``important'' than the segment selector,
which provides just a ``context'' for the translation.}
\code{ea-to-la} takes as inputs the machine state and a logical address,
which consists of an effective address and (the index of) a segment register;
it returns as output a linear address,
obtained by adding the effective address to the base of the segment.
\code{ea-to-la} also returns as output a flag with error information
if the effective address is outside the segment's bounds;
error flag and linear address are returned via \code{mv}.
The segment base and bounds are retrieved
via the function \code{segment-base-and-bounds} described earlier.
In 64-bit mode, there are no checks against the segment's bounds,
but instead the resulting address is checked to be canonical;
the resulting address differs from the input effective address
only if the segment is \code{FS} or \code{GS},
whose base is added to the effective address.

The 64-bit model included top-level memory access functions
to read and write unsigned and signed values of different sizes
(bytes, words, double words, etc.).
These were called \code{rmXX}, \code{rimXX}, \code{wmXX}, and \code{wimXX},
where \code{r} indicates `read',
\code{w} indicates `write',
\code{i} indicates `integer'
(vs.\ natural number,
i.e., the functions with \code{i} read/write signed integers
while the functions without \code{i} read/write unsigned integers),
and \code{XX} consists of two digits indicating the bit size
(\code{08}, \code{16}, \code{32}, etc.).
There were also more general functions
\code{rm-size}, \code{rim-size}, \code{wm-size}, and \code{wim-size},
which took the size in bytes as an additional input.
These functions took linear addresses as inputs;
they assumed
that the base of the \code{FS} or \code{GS} segment
had already been added when applicable,
and that alignment checks had already been performed when needed.

New top-level memory access functions were added
to read and write unsigned and signed values of different sizes,
mirroring the aforementioned ones.
They are called
\code{rmeXX}, \code{rimeXX}, \code{wmeXX}, \code{wimeXX},
\code{rme-size}, \code{rime-size}, \code{wme-size}, and \code{wime-size},
where \code{e} stands for `effective address'.
These new functions
take as inputs logical addresses
(as effective addresses and segment registers)%
\footnote{A similar observation applies to these names,
as the one made for the name of \code{ea-to-la} above.}
instead of linear addresses,
call \code{ea-to-la} to obtain linear addresses,
perform alignment checks on the linear addresses
if indicated by some additional inputs to these new functions,
and then call the old 64-bit mode top-level memory access functions
(which are no longer at the top level in the extended model).
For clarity, the old functions were renamed to
\code{rmlXX}, \code{rimlXX}, \code{wmlXX}, \code{wimlXX},
\code{rml-size}, \code{riml-size}, \code{wml-size}, and \code{wiml-size},
where \code{l} stands for `linear address'.

The inputs of all these functions (new and old) include the machine state;
the inputs of the write functions also include the value to write,
while the inputs of the read functions also include
a flag \code{:r} or \code{:x}
to distinguish between reading data and fetching instructions.
The read functions return
an error flag, the value read, and the possibly updated machine state
(see Footnote~\ref{footnote:read-may-change-state}),
via \code{mv}.
The write functions return
an error flag and the possibly updated machine state, via \code{mv}.

\subsection{Paging}\label{sec:extension:paging}

As explained in \secref{sec:background-x86},
there are three kinds of paging in the x86 processor.
PAE and 32-bit paging are used only in 32-bit mode, and 4-level paging is used only in 64-bit mode.
The 64-bit model included the latter, but not the former.%
\footnote{To be precise, a file in the 64-bit model contained
a start towards modeling the two paging modes used in 32-bit mode,
but this file's content was not used in the rest of the model.}

As explained in \secref{sec:background-model},
the 64-bit model included the specification of paging only in its
system-level view, not in its application-level view.  These views
carry over to the extended model as well.
However, the extended model supports 32-bit mode only in application-level view for now,
that is, it only supports the execution and verification
of 32-bit application software, not 32-bit system software.
As such, we have deferred the addition of PAE and 32-bit paging to the model
(see \secref{sec:future}).

\subsection{Instruction Fetching}\label{sec:extension:fetching}

The 64-bit model fetched instruction bytes
by using the instruction pointer in the \code{RIP} register
as the linear address passed to
the top-level memory access functions \code{rm08} etc.\
mentioned in \secref{sec:extension:segmentation}.
The instruction pointer was incremented using ACL2's \code{+} operation,
checking that the result was canonical after each increment.

In 32-bit mode,
the instruction pointer is in the \code{EIP} or \code{IP} register,
which refer to the low 32-bit or 16-bit portions of the \code{RIP} register, respectively.
The \code{D} bit in the \code{CS} register selects the appropriate register:
\code{EIP} if this bit is set and \code{IP}
otherwise.

A new function \code{read-*ip}
(where the \code{*} is meant to stand for \code{r}, \code{e}, or nothing)
was added, to generalize the 64-bit mode reading
of the instruction pointer from the \code{RIP} register.
This new function takes as input the machine state
and returns as output the instruction pointer.
In 32-bit mode, the output is a 32-bit or 16-bit value,
from the low bits of the \code{rip} component of the model of the x86 state.

A new function \code{add-to-*ip} was added
to generalize the plain 64-bit increment of the instruction pointer
and the subsequent canonical address check.
This new function takes as inputs
the machine state,
an instruction pointer,
and an integer amount to add (which may be negative);
it returns as outputs the incremented instruction pointer
and an error flag (\code{nil} if there is no error).
In 32-bit mode,
the input and output instruction pointers are 32-bit or 16-bit values,
and the error flag is non-\code{nil}
if the output instruction pointer
is outside the limits of the \code{CS} register.
Note that the input machine state is only used to check the \code{CS} limits,
and that there is no output machine state:
this function only increments instruction pointer values.

A new function \code{write-*ip} was added
to store the final instruction pointer, after executing each instruction,
into the \code{RIP}, \code{EIP}, or \code{IP} register.
This function takes as inputs
the machine state
and the instruction pointer to store;
it returns as output an updated machine state.
In 32-bit mode,
the instruction pointer is a 32-bit or 16-bit value
that is stored in the low 32 or 16 bits of
the \code{rip} component of the model of the x86 state.

The model of instruction fetching was modified to call the new functions
\code{read-*ip} and \code{add-to-*ip} to manipulate the instruction pointer
(while \code{write-*ip} is called by the models of the individual instructions),
and to replace the calls to \code{rml08}
(i.e., the renamed \code{rm08}; see \secref{sec:extension:segmentation})
with calls to \code{rme08}.
This change necessitated the largest effort,
compared to the other changes to the 64-bit model,
to keep all the proofs working:
see \secref{sec:extension:proofs} for details.

\subsection{Stack Operations}\label{sec:extension:stack}

Several instructions read from and write to the stack, via the stack pointer.
The 64-bit model manipulated the stack pointer
in a manner similar to the instruction pointer
(see \secref{sec:extension:fetching}):
the stack pointer was read from and written to the \code{RSP} register,
and it was incremented and decremented
via ACL2's \code{+} and \code{-} operations,
checking that the result was canonical after each increment and decrement.

In 32-bit mode, the stack pointer is in the \code{ESP} or \code{SP} register,
which are the low 32-bit or 16-bit portions of the \code{RSP} register, respectively.
The \code{B} bit in the \code{SS} register selects the appropriate register:
\code{ESP} if this bit is set and \code{SP}
otherwise.

The extensions to the model for the stack pointer
were similar to the extensions for the instruction pointer.
New functions \code{read-*sp}, \code{add-to-*sp}, and \code{write-*sp}
were added,
quite analogous to the functions
\code{read-*ip}, \code{add-to-*ip}, and \code{write-*ip}
described in \secref{sec:extension:fetching}.
Instead of \code{rip}, \code{read-*sp} and \code{write-*sp} read and write
the \code{rsp} component of the model of the x86 state ---
only the low 32 or 16 bits, in 32-bit mode.
Instead of the limits of the \code{CS} segment,
\code{add-to-*sp} checks the new stack pointer value against
the limits of the \code{SS} segment,
taking into account the possibility of expand-down segments,
which are typically used for the stack.

\subsection{Addressing Modes}\label{sec:extension:addressing}

Other than stack and immediate operands,
instructions can reference operands in memory
via a variety of addressing modes.
Each addressing mode involves
a calculation of an effective address from one or more components,
such as (the content of) a base register,
an index to add to the base register,
and a scaling factor with which to multiply the index
before adding it to the base.
The addressing mode to use is specified by
the \code{ModR/M}, \code{SIB}, and displacement bytes of an instruction
(not all of these bytes need to be present).
There are 32-bit and 16-bit addressing
modes~\cite[Tables 2.1 and 2.2]{intel-manuals}:
the former apply to 64-bit mode,
and to 32-bit mode when the address size is 32 bits;
the latter apply to 32-bit mode when the address size is 16 bits.
The address size is determined by the \code{D} bit of the \code{CS} register.

The 64-bit model included functions
to decode the \code{ModR/M} and \code{SIB} bytes
and to perform the effective address calculations
for the 32-bit addressing modes,
but not for the 16-bit addressing modes, which do not apply to 64-bit mode.
In particular, the function \code{x86-effective-addr}
performed the \code{ModR/M} and \code{SIB} decoding,
and the effective address calculation,
for the 32-bit addressing modes.
Since the 16-bit addressing modes are fairly different from
the 32-bit addressing modes,
\code{x86-effective-addr} was renamed to \code{x86-effective-addr-32/64}
(to convey that this is used for 32-bit and 64-bit addresses),
a new function \code{x86-effective-addr-16} was added
for the 16-bit addressing modes,
and a new wrapper function \code{x86-effective-addr} was added that calls
either \code{x86-effective-addr-32/64} or \code{x86-effective-addr-16}.

These functions have several inputs and several outputs.
The inputs include the \code{ModR/M} and \code{SIB} bytes (0 if absent),
the current instruction pointer
(which is just past the opcode and,
if present, the \code{ModR/M} and \code{SIB} bytes),
and the machine state.
The outputs are an error flag,
the calculated effective address,
the number of (displacement) bytes read
(which is later added to the instruction pointer),
and a possibly updated machine state.
Based on the addressing mode,
these functions may read the displacement bytes,
advancing the instruction pointer;
they call \code{rme08} to read these bytes,
which may fail (see \secref{sec:extension:segmentation}),
in which case \code{x86-effective-addr} returns a non-\code{nil} error flag.

\subsection{Operand Reading and Writing}\label{sec:extension:operands}

Some instructions read and write their operands
from and to memory or registers,
based on some bits of the \code{ModR/M} byte ---
which, as explained in \secref{sec:extension:addressing},
is also used to determine the addressing mode when the operand is in the memory.

The 64-bit model included a function
\code{x86-operand-from-modr/m-and-sib-bytes}
that uniformly read an operand from memory or registers
based on the \code{ModR/M} and \code{SIB} bytes.
This function decoded the \code{ModR/M} byte,
and, based on that, either read the operand value from a register,
or from memory.
In the latter case,
it called \code{x86-effective-addr} to calculate the effective address,
which it passed to
\code{rml-size} (formerly \code{rm-size}) or its variants
to read the value from that location;
recall that in the 64-bit model an effective address
was also a linear address.
This function had several inputs and several outputs.
The inputs included the \code{ModR/M} and \code{SIB} bytes,
the current instruction pointer
(which is just past the opcode and,
if present, the \code{ModR/M} and \code{SIB} bytes),
and the machine state.
The outputs included an error flag,
the memory or register operand read,
the calculated effective address (for a memory operand),
the number of (displacement) bytes read
(which is later added to the instruction pointer),
and a possibly updated machine state.
The calculated effective address is returned
to avoid recalculating it when an instruction updates the operand,
which is common (e.g., adding an immediate value to an operand in memory).

To uniformly write operands to memory or registers, the 64-bit model also included functions
\code{x86\-operand-to-reg/mem} and \code{x86-operand-to-xmm/mem}.
These functions had several inputs,
including the value to write,
the effective/linear address
(calculated externally,
often by
\code{x86-operand-from-modr/m\-and-sib-bytes} as mentioned above),
and the machine state.
These functions returned an error flag and an updated machine state.
The effective/linear address was passed to
\code{wml-size} (formerly \code{wm-size}) or its variants
to write the operand value to memory.

To support 32-bit mode,
these operand read and write functions had to be generalized
to traffic in effective addresses rather than linear addresses.
In particular, the index of the segment register to use
had to be an additional input.
This required a change to the interface of these three functions,
which are called in many places by the instruction models.
Making this change in one shot
would have required a lot of adjustments to instruction models
and could have broken many proofs.
Thus, the change was staged as follows:
(1) introduce new versions of these functions,
called \code{x86-operand-from-modr/m-and\-sib-bytes\$},
\code{x86-operand-to-reg/mem\$}, and \code{x86-operand-to-xmm/mem\$},
with the new interface and functionality;
(2) separately modify the instruction models, one by one,
to call the new functions instead of the old ones,
repairing any failing proofs;
(3) remove the old functions when they are no longer called;
(4) rename the new functions to remove the ending \code{\$}.
The new functions to read and write operands
call the new top-level memory functions
\code{rme-size} and \code{wme-size} or their variants
to access the operands in memory.

The segment register to pass to the new functions is determined using
the default segment selection rules~\cite[Volume 1, Table 3-5]{intel-manuals}
based on the kind of instruction and the addressing mode,
and the presence of a segment-override prefix before the opcode.
This determination is made by a new function \code{select-segment-register}
that was added for this purpose.
This function has several inputs,
including the segment override prefix (0 if absent)
and the machine state;
this functions returns (the index of) a segment register as the only output.

\subsection{Instructions}

With all the extensions described above in place,
extending the individual instructions' semantic functions to 32-bit mode
was comparatively easy.
This was also facilitated by the fact that
the 64-bit model already supported different operand sizes
for many of the core operations carried out by the instructions
(e.g., the arithmetic and logic operations)
leaving just the operand size determination logic of the instruction
to be extended.

In particular, the existing function \code{select-operand-size}
was extended to return the operand size in 32-bit mode,
while returning the same result as before in 64-bit mode.
The interface of this function was expanded
to give it access to the \code{D} bit of the \code{CS} register,
needed to determine the operand size in 32-bit mode:
its inputs include
the machine state and some of the decoded instruction prefixes;
it returns the size as the only output.
The instruction semantic functions
that already called \code{select-operand-size}
could then automatically extend their operand size determination to 32-bit mode.

A new function \code{select-address-size} was introduced
to determine the address size in both 32-bit and 64-bit mode.
Its inputs include
the machine state and some of the decoded instruction prefixes;
it returns the size as the only output.
The portions of the instruction semantic functions
that determined the address size in 64-bit mode only
were replaced with calls to this new function.

In the 64-bit model, the instruction semantic functions were
reading immediate operands via \code{rm08} and related functions
(renamed to \code{rml08} etc.)
called on the instruction pointer (treated as a linear address),
incrementing the instruction pointer via ACL2's \code{+},
and writing the final instruction pointer
directly into the \code{rip} component of the x86 state.
To support 32-bit mode as well, this was changed to
read immediate operands via \code{rme08} and related functions
called on the instruction pointer
(treated as an effective address in the code segment),
and to use the new functions described in \secref{sec:extension:fetching}
to increment and store the instruction pointer.

In the 64-bit model, the instruction semantic functions were
reading and writing the stack pointer directly
from and to the \code{rsp} component of the x86 state,
calling \code{rm-size}, \code{wm-size}, and related functions
(renamed to \code{rml-size}, \code{wml-size}, etc.)
on the stack pointer (treated as a linear address)
to read and write stack operands, and
incrementing and decrementing the stack pointer
via ACL2's \code{+} and \code{-}.
To support 32-bit mode as well, this was changed to
call \code{rme-size}, \code{wme-size}, and related functions
on the stack pointer (treated as an effective address in the stack segment)
to read and write stack operands,
and to use the new functions described in \secref{sec:extension:stack}
to read, write, increment, and decrement the stack pointer.
The instruction semantic functions in the 64-bit model
were performing alignment checks on the stack pointer as needed:
this code was removed because alignment checks
are performed inside \code{rme-size}, \code{wme-size}, and related functions,
after effective addresses are translated to linear addresses,
resulting in better factored code as a by-product.

Following the staged approach described in \secref{sec:extension:operands},
calls to \code{x86-operand-from-modr/m-and\-sib-bytes} etc.\
in the instruction semantic functions,
were gradually redirected to call
\code{x86-operand\-from-modr/m-and-sib-bytes\$} etc.\ instead.
Since these new functions traffic in effective addresses
instead of linear addresses,
two additional adjustments had to be made to the instruction semantic functions.
The first adjustment was to remove alignment checks
performed on the linear addresses of the operands:
alignment checks are already performed by
(functions called by) the new functions,
after translating the now effective addresses of the operands
into linear addresses.
The second adjustment was to remove
the addition of the \code{FS} or \code{GS} segment base (when applicable)
to the linear addresses of the operands:
any segment base is added to the now effective addresses of the operands
when they are translated into linear addresses,
by (functions called by) the new functions.
These adjustments resulted in better factored code as a by-product.

After each instruction semantic function was extended to 32-bit mode
in the manner explained above,
the top-level dispatch was adjusted to call the function
not only in 64-bit mode, but also in 32-bit mode.
Concretely, this was done by removing the wrapping conditional
\code{(if (64-bit-modep x86) ...)}
described in \secref{sec:extension:mode}.

All the non-floating-point instructions of the 64-bit model
have been extended to 32-bit mode,
at least for the application-level view of execution.%
\footnote{The far variant of \code{JMP} does not handle
32-bit mode system segments yet,
but these are only needed for the system-level view of execution.
Nonetheless, we plan to add support for these soon.}
The 34 floating-point instructions supported by the 64-bit model
also need to be extended to 32-bit mode. We anticipate that this task
will be straightforward because we have almost all the pieces
necessary to extend the floating-point instructions.  Specification
functions from Russinoff's RTL library~\cite{acl2:books:rtl-rel11} are
used to specify the core operations of these instructions, and these
functions are already parameterized by size.  As far as operand access
and update is concerned: we have already extended memory access
functions to work in 32-bit mode, but since the floating-point
instructions use a different register set from the general-purpose
instructions, we just need to extend register access functions to work
in 32-bit mode.

\subsection{Proof Adaptations}\label{sec:extension:proofs}

In extending the 64-bit ISA model to support 32-bit mode, so far we
have discussed issues that were not especially unique to a formal
model; for instance, extending a C emulator of the 64-bit x86 ISA to
support the execution of 32-bit programs would likely require making
similar changes.  However, as discussed in
\secref{sec:extension:challenges}, our aim was also to preserve the
proofs done in the 64-bit model, including the proofs of the model's functions
(e.g., guards)
and the proofs of correctness of the 64-bit programs.

The proofs of 64-bit programs with the 64-bit model
involved functions and expressions
that were generalized for 32-bit mode as described in the previous subsections.
Thus, theorems were added to rewrite the more general functions and expressions
into the ``old'' ones under the \code{64-bit-modep} hypothesis.
For instance, the following theorem was added,
to rewrite the reading of the \code{RSP}, \code{ESP}, or \code{SP} register
into just the reading of \code{RSP}
(where \code{rgfi} reads the 64-bit value of the register specified as argument):
\begin{codeblock}
  (implies (64-bit-modep x86)
           (equal (read-*sp x86) (rgfi *rsp* x86)))
\end{codeblock}
Since the 64-bit proofs include the \code{64-bit-modep} hypothesis
(added as described in \secref{sec:extension:mode}),
these rewrite rules fire and help reduce (sub)goals to the form
they had in the 64-bit model prior to its extensions.
Additional theorems had to be added in order to let the
\code{64-bit-modep} hypothesis be relieved for updated x86 states,
such as the following one,
asserting that reading from a linear address does not change mode:%
\footnote{\label{footnote:read-may-change-state}%
Note that reading from memory may change the x86 state in general,
by changing the `accessed' flag in the paging data structures;
this flag is discussed later in this section.
\code{rml08} and similar functions return the possibly updated state
as the third component of their multi-value result,
accessed via \code{(mv-nth 2 ...)}.}
\begin{codeblock}
  (equal (64-bit-modep (mv-nth 2 (rml08 addr r-x x86)))
         (64-bit-modep x86))
\end{codeblock}
All these theorems generally sufficed to keep the proofs of 64-bit application
programs, i.e., in the application-level view.

Even though the extended model does not (yet) cover the system-level view,
additional changes were needed to adapt the reasoning strategy to work
for the proofs of some 64-bit system programs, as described below.
Lemma libraries included in the 64-bit model for
formally analyzing system programs contained utilities to reason about
the memory management data structures --- specifically, the paging
structures.  Every linear-to-physical address translation on the x86
ISA causes a traversal of entries in these paging
structures,\footnote{We ignore translation look-aside buffers and
  caches for this discussion.} which produces some side effects.  The
processor can set two bits in the paging entries during address
translation: the {\em accessed} and {\em dirty} flags, which
effectively {\em mark} the entries that govern the translation of a
linear address.  However, the mapping of that linear address to its
corresponding physical address does not change as a result of these
side-effect updates.
This means that statements like the following are true, where
\code{spec-function} is an x86 specification function that reads from
or writes to memory, and {\code{<hyp>}} ensures that (1)
\code{x86-1} and \code{x86-2} can differ only in the values of the
accessed and dirty flags, and (2) \code{spec-function} traffics only
in locations that are disjoint from the affected (i.e., marked) entries in the paging
structures:
\begin{codeblock}
  (implies <hyp> (equal (spec-function <args> x86-1) (spec-function <args> x86-2)))
\end{codeblock}
The 64-bit model stored such theorems as congruence rules, as follows:
\begin{codeblock}
  (implies (xlate-equiv-memory x86-1 x86-2)
           (equal (spec-function-alt <args> x86-1)
                  (spec-function-alt <args> x86-2)))
\end{codeblock}
The function \code{spec-function-alt} is exactly the same as
\code{spec-function} under one condition:
it traffics only in locations
that are disjoint from all the paging entries that are marked during address translation; if this condition
is false, then \code{spec-function-alt} simply returns the input x86
state --- unmodified.  The relation \code{xlate-equiv-memory} says that
two x86 states are equivalent if: (1) their paging structures are
equal, modulo the accessed and dirty bits; and (2) all other memory
locations are equal.  This arrangement facilitates ``conditional''
congruence-based reasoning\footnote{Recall that a congruence rule
  cannot have anything other than a known equivalence relation as its only
  hypothesis.} --- a rewrite rule transforms \code{spec-function} to
\code{spec-function-alt} whenever applicable, and then these
congruence rules can come into play.  This strategy works well for
system programs that do not explicitly modify the paging data
structures.
As a result of extending the model, we had to add \code{64-bit-modep}
both to \code{xlate-equiv-memory} and to the conditions under which
\code{spec-function-alt} is equal to \code{spec-function}.

We note that the top-level rewrite rules in the model are not
mode-specific --- an example is the opener lemma of the step function,
whose hypotheses determine when to rewrite a call of the step function
to the function that dispatches control to the appropriate instruction
semantic function; basically, this lemma helps in ``unwinding'' the
x86 interpreter during proofs.  Thus, we share lemmas across different
modes as much as possible.  Rules about intermediate specification
functions that do have a hypothesis like \code{(64-bit-modep x86)}
come in useful during program verification when we need to be
cognizant of the operating mode of a processor.

\subsection{Performance}

As discussed in Section~\ref{sec:background-model}, a high simulation
speed is crucial for performing model validation via co-simulations
against real machines.  We first discuss how the simulation speed of
64-bit programs was adversely impacted by extending the model to
support 32-bit programs and how we fixed this issue.  Then we present
performance-related information about the simulation of 32-bit programs.

\subsubsection{64-bit Mode}

Our initial attempt to extend the 64-bit model led to an unfortunate
decrease in its simulation speed in the application-level view from
around 3 million instructions/second to 1.9 million
instructions/second.\footnote{There was a similar decrease in speed in
  the system-level view, but for this discussion, we focus only on the
  application-level view.}  Using Linux's Perf utility and ACL2's
\code{profile} mechanism
\acldoc{profile}{http://www.cs.utexas.edu/users/moore/acl2/manuals/current/manual/?topic=ACL2____PROFILE}, we discovered that the predicate
\code{64-bit-modep} was the main culprit --- every call of this
function allocated 16 bytes on the heap.  Recall that this predicate
reads values from fields in the x86 state (\code{IA32\_EFER} and
\code{CS} registers) and these fields usually contain large integers,
i.e., {\em bignums}.  Manipulating bignums is inefficient because they
are stored on the heap and require special arithmetic functions.
Contrast these bignums with {\em fixnum} integers, which can fit in
the host machine's registers and whose computation can be done with
built-in arithmetic instructions.  For example, Lisp compilers can add
two fixnum values using the add instruction of the host machine, but
the addition of two bignum values is done by calling a Lisp function
that is considerably slower than a machine instruction and that may
also allocate memory on the heap.

Though \code{64-bit-modep} allocates just 16 bytes on the heap, it is
called once at the beginning of the execution of {\em each}
instruction to determine the processor's operating mode
for that instruction.
Therefore, the number of bytes allocated because of this function
 increases with the number of instructions to be executed.
For instance, for a simple C function that computes the \code{n}-th
Fibonacci number by implementing the Fibonacci recurrence relation,
the number of 64-bit x86 instructions to be executed to compute
\code{fib(30)} was 31,281,993.  This program run allocated 500,511,936
bytes on the model, all because of the bignum computations done in
\code{64-bit-modep}.  Since we typically run a large number of
instructions at once during co-simulations, this problem needed to be
fixed.

We solved this issue by changing the definition of \code{64-bit-modep}
--- it is now an {\tt mbe}, where our initial definition has been
retained in the \code{:logic} part.  For the \code{:exec} part, we
used a function called \code{bignum-extract} from a pre-existing ACL2
community book
\acldoc{bignum-extract}{http://www.cs.utexas.edu/users/moore/acl2/manuals/current/manual/?topic=BITSETS____BIGNUM-EXTRACT}.  Logically,
\code{(bignum-extract x n)} returns the \code{n}-th 32-bit slice of an
integer \code{x}.  Therefore, in the \code{:exec} part of
\code{64-bit-modep}, we simply extracted relevant 32-bit slices of the
x86 fields using this function and read the required values from them.
During execution, we used a raw Lisp replacement for
\code{bignum-extract} (already provided in
\code{std/bitsets/bignum-extract-opt.lisp}) that takes advantage of
64-bit CCL's implementation.  CCL stores bignums in vectors of 32
bits, and extracting a 32-bit chunk of a bignum to operate on avoids
expensive computations involving the {\em entire} bignum.

With this solution, the model's simulation speed for 64-bit programs went back
to what it was at the beginning of this project (for both
application-level and system-level views).  We note that the number of
bytes allocated on the heap to compute \code{fib(n)}
is now independent of the number of instructions to be executed.

However, this solution does have two drawbacks:
\begin{enumerate}[nosep]
  \item The raw Lisp replacement for \code{bignum-extract} may not be
    efficient for Lisps other than CCL.
  \item For efficient execution (not for reasoning), this solution
    requires a trust tag.
\end{enumerate}

In the future, we plan to implement a different solution that will not
suffer from the above drawbacks.  We will take advantage of the x86
abstract stobj by defining the fields involved in the computation of
\code{64-bit-modep} as multiple fixnum-sized fields in the concrete
stobj, defining accessor and updater functions over the concrete stobj
that operate on these fields collectively, and then proving that these
concrete functions correspond to the currently existing accessor and
updater functions defined over the abstract stobj.  A benefit of this
approach is that after the abstract stobj is defined, no other proofs
would be affected.

\subsubsection{32-bit Mode}

The simulation speed of 32-bit programs in the application-level view
is around 920,000
instructions/ second --- note that, for 32-bit mode, the
system-level view has not been implemented yet.  One reason for this
slower speed (as compared to 64-bit mode) is that 32-bit programs
often have to refer to segment descriptors over the course of their
execution, and these operations usually involve bignums.

Our focus has been on the functionality of the extensions first rather
than the performance, but in the future, we intend to speed up
32-bit mode using the same kind of approaches as those used in
64-bit mode.  That being said, the current simulation speed of
32-bit programs has been more than adequate to run tests at this stage of
development.

\subsection{Documentation}

As mentioned in \secref{sec:background-model},
the 64-bit model was extensively documented.
As the model was being extended to support 32-bit mode,
the documentation was extended accordingly.


\section{Related Work}\label{sec:related}

The x86-CC and x86-TSO models of the x86 ISA in HOL~\cite{x86popl,tso10}
are focused on concurrent memory access in multiprocessor systems.
They support a few tens of instructions,
defined via parameterized monadic ``micro-code'' operations
that can be instantiated for both sequential and concurrent semantics.
The models are accompanied by
a tool to test the sequential semantics in HOL, and
a tool to test the concurrent semantics in (unverified) OCaml.

The model of the x86 ISA in Coq developed for RockSalt~\cite{rocksalt}
supports the verification of a binary analyzer
that checks conformance to a sandboxing policy;
however, the model is independent of this analyzer.
It supports about 100 instructions in 32-bit mode only,
and does not model paging tables.
It uses two Domain-Specific Languages (DSLs) embedded in Coq to declaratively specify instruction decoding
and to express instruction semantics in terms of ``micro-code'' operations.
The model can simulate and validate (against real processors)
about 50 instructions per second.

Degenbaev's model of the x86 ISA~\cite{degenbaev2012formal}
supports both concurrent memory access in multiprocessor systems
and a large number of instructions; in this work, the instruction semantics is modeled via a DSL.
Baumann's related model of the x87 instruction set~\cite{baumann2008formal}
supports floating-point instructions.
These models are written in ``pencil-and-paper'',
not in a theorem prover or other formal tool.

Models of the ISA of other processors have been developed
in ACL2 and other theorem provers,
e.g.,~\cite{sh:pipeline,armv7,reidtrustworthy}
In particular,
the model of the y86 ISA (an x86-like ISA) in ACL2~\cite{acl2:books:y86}
is a precursor of our model.
The modeling approaches for these other processors are similar to
the ones for the x86 ISA models in ACL2 and other provers,
but the ISAs of these other processors are less complicated than the x86.

Models of both intermediate-level and source-level languages
have been also developed in ACL2 and other theorem provers,
e.g.,~\cite{liu2004java,norrish-c}.
There are some commonalities
between the modeling approaches for these languages
and the modeling approaches for processor ISAs,
such as the suitability of an interpreter style of operational semantics.
There are also many differences,
e.g., processor ISAs typically have fewer data types and do not have
to include specifications of operations like casting (e.g., converting
integers to characters, using pointers as integers, etc.).

There exist several x86 emulators (e.g., Bochs, DOSBox, PCem, QEMU, Unicorn)
that are written in more conventional programming languages (e.g., C++),
not in the logical languages of theorem provers.
The development of these emulators faces many of the same issues
as the development of a formal model,
such as the sheer complexity and the occasional ambiguity
of the official documentation.
However, the proof aspects are unique to formal models.
In particular,
extending or improving an emulator involves regression testing,
while extending or improving a formal model may also involve proof adaptation.


\section{Future Work}\label{sec:future}

In this paper, we discussed our ongoing effort to extend the
pre-existing 64-bit model of the x86 ISA to support Intel's 32-bit mode.
In the short or medium term, we plan to add support for the following
x86 ISA features:
\begin{itemize}[nosep]
\item
More instructions,
prioritizing on the ones encountered in new programs to verify
and on the ones that are more commonly used in general.
Every new instruction will include both 64-bit and 32-bit mode support.
\item
The two kinds of paging used in 32-bit mode (see \secref{sec:extension:paging}).
This will enable the formal analysis of 32-bit system software,
besides 32-bit application software.
\item
The remaining processor modes
(real-address, virtual-8086, and system management),
along with the instructions to switch between modes.
\item
The widely-used Intel's vector processing features
(i.e., AVX, AVX2, AVX-512) that offer enhanced capabilities to perform
data-intensive (integer and floating-point) computations efficiently.
The x86 model already supports the decoding of these instructions, but 
this task will entail modifying some old
instructions that now support these new features and adding many new
instructions.  A notable thing is that vector
processing features differ between the 64-bit and 32-bit modes,
thereby making this task a little more interesting.
\end{itemize}
In the short or medium term, we also plan to validate the 32-bit extensions
against actual x86 processors, as was done for the 64-bit model.

A longer-term project is to support a concurrent semantics.
At the very least, this will likely require some refactoring of the instruction semantic functions,
which currently treat the execution of each instruction as an atomic event,
because the x86 state is updated at the end of each instruction.

Another direction is to make some parts of the specification
more declarative and concise via more macros to generate boilerplate code.
Examples are
the binding of various instruction bytes and fields (e.g., prefixes) to variables
and common checks
(e.g., many instructions throw a \code{\#UD} exception
if the \code{LOCK} prefix is used).
A more ambitious endeavor is
to raise the level of abstraction of some parts of the specification
to the point of making them non-executable,
and then use transformation tools like APT~\cite{apt-www,apt-simplify}
to generate efficient executable refinements of the specification
along with correctness proofs checked by ACL2.

As briefly mentioned in \secref{sec:introduction},
some work has started on 32-bit program verification,
mainly aimed at detecting malware variants via semantic equivalence checking
by symbolic execution.
This work,
performed by Eric Smith with some help about the model from the first author,
will be carried forward in the near future.

Longer-term envisioned uses of the model include
verified compilation from programming languages (e.g.\ C) to binaries,
as well as binary program derivations by stepwise refinement
using transformation tools like APT~\cite{apt-www,apt-simplify},
possibly on deeply embedded representations~\cite{coglio-pop}.

\section*{Acknowledgements}
This work was partially supported by DARPA under Contract No.\ D16PC00178.
Thanks to Eric Smith and James McDonald for providing feedback
based on their use of the model for proofs, symbolic execution, and simulation.

\vspace{-2mm}
\bibliographystyle{eptcs}
\bibliography{top}

\end{document}